\documentclass[12pt]{article} 

\pdfoutput=1

\usepackage[]{amsmath}
\usepackage[]{graphicx}
\usepackage[]{latexsym}
\usepackage{geometry}
\newcommand{\reef}[1]{(\ref{#1})}
\def\beq{\begin{equation}} 
\def\eeq{\end{equation}} 
\def\barr{\begin{array}} 
\def\earr{\end{array}} 
\def\beqa{\begin{eqnarray*}} 
\def\eeqa{\end{eqnarray*}} 
 
\font\mybb=msbm10 at 12pt 
\def\bb#1{\hbox{\mybb#1}}

\def\b1 {\bb{1}} 

\def\half {\frac{1}{2}} 

\def\lesssim{\mathrel{\hbox{\rlap{\hbox{\lower4pt\hbox{$\sim$}}}\hbox{$<$}}}} 
\def\gtrsim{\mathrel{\hbox{\rlap{\hbox{\lower4pt\hbox{$\sim$}}}\hbox{$>$}}}}

\begin{document} 

\vspace*{-.6in} \thispagestyle{empty}
\begin{flushright}
\end{flushright}
\vspace{.2in} {\Large
\begin{center}
{\bf On three-point  correlation functions in the gauge/gravity duality}
\end{center}}
\vspace{.2in}
\begin{center}
Miguel S. Costa$^{a}$, Ricardo Monteiro$^{a,b}$, Jorge E. Santos$^{a,b}$, Dimitrios Zoakos$^{a}$ 
\\
\vspace{.3in} 
\emph{${}^a\,$
Centro de F\'\i sica do Porto\\
Departamento de F\'\i sica e Astronomia\\
Faculdade de Ci\^encias da Universidade do Porto\\
Rua do Campo Alegre 687,
4169--007 Porto, Portugal}
\\
\vspace{.2in} 
\emph{${}^b\,$DAMTP, Centre for Mathematical Sciences \\ University of Cambridge\\
Wilberforce Road, Cambridge CB3 0WA, UK}

\end{center}

\vspace{.3in}

\begin{abstract}
We study the effect of marginal and irrelevant deformations on 
the renormalization of operators near a CFT fixed point.
New divergences in a given operator are determined by its  OPE with the operator ${\cal D}$ that generates the deformation.
This provides a scheme to compute the couplings $a_{{\cal D}AB}$ between the operator ${\cal D}$ and
two arbitrary operators ${\cal O}_A$ and ${\cal O}_B$. We exemplify for the case of ${\cal N}=4$ SYM,
considering the simplest case of the exact  Lagrangian deformation. 
In this case the deformed anomalous dimension matrix is determined by the derivative 
of the anomalous dimension matrix with respect to the coupling. 
We use integrability techniques to compute the one-loop couplings $a_{{\cal L}AB}$
between the Lagrangian and two distinct large operators built with Magnons, 
in the SU(2) sector of the theory.
Then we consider  $a_{{\cal D}AA}$ at
strong coupling, and show how to compute it using the gauge/gravity duality, 
when  ${\cal D}$  is a chiral operator dual to any supergravity field and 
${\cal O}_A$ is dual to a heavy string state. We exemplify for the Lagrangian and operators  ${\cal O}_A$ 
dual to  heavy string states, showing agreement with the
prediction derived from the renormalization group arguments.
\end{abstract}

\newpage

\section{Introduction}
To solve a Conformal Field Theory (CFT) amounts to finding its spectrum and 3-point correlation
functions, since higher point functions may be obtained using the operator product expansion (OPE). 
In the former case this means finding the anomalous dimensions of  the operators of the 
theory, while in the latter case it means finding the couplings in 3-point correlation functions, whose
space-time dependence is otherwise fixed by conformal invariance. In the simplest case of
scalar primary operators the 3-point function has the simple form
\beq
\left\langle {\cal O}_A(0){\cal O}_B(x){\cal O}_C(y)\right\rangle =
\frac{a_{ABC}}{|x|^{\Delta_A+\Delta_B-\Delta_C}|y|^{\Delta_A+\Delta_C-\Delta_B}|x-y|^{\Delta_B+\Delta_C-\Delta_A}}\,.
\eeq
where $\Delta_A$ is the dimension of the operator ${\cal O}_A$, and so on. The definition of the couplings $a_{ABC}$
requires that the operators diagonalise the anomalous dimension matrix and depends on the choice of normalisation in the 
2-point function of each operator. 

Our main interest is to explore new methods to compute the couplings $a_{ABC}$ for certain single trace operators in
${\cal N}=4$ SYM. In recent years there have been great progresses in
finding the spectrum of this theory, in the planar limit and 
for any value of the coupling constant, using integrability
\cite{Bena:2003wd}-\cite{Bombardelli:2009ns}. On the other hand, much remains to be done in the computation of the
couplings $a_{ABC}$. At weak coupling these may be evaluated, order by order in perturbation theory,  by computing Feynman diagrams
\cite{Kristjansen:2002bb}-\cite{Grossardt:2010xq}. 
Although this approach is essential to uncover new structures and to verify  new exact results, it is unpractical to obtain exact results for 
general operators. A more promising approach is to explore integrability of planar ${\cal N}=4$ SYM. However, how integrability 
will enter computations of the couplings $a_{ABC}$ remains unclear. 

One strategy to compute the couplings in a CFT is to deform the theory from its fixed  point with a marginal or irrelevant operator ${\cal D}$. 
We will show in Section 2 that this deformation introduces new divergences in the renormalised operators of the critical theory, which are 
determined by the couplings $a_{{\cal D}AB}$. More precisely, to leading order in the deformation parameter,  the entry of
the deformed anomalous dimension matrix between operators ${\cal O}_A$ and ${\cal O}_B$ is determined by the coupling $a_{{\cal D}AB}$. 
Thus, in planar ${\cal N}=4$ SYM,
finding the action of such matrix on operators diagonalized by means of the Bethe ansatz is a new method to compute the couplings $a_{{\cal D}AB}$.
In practice, we will show
in Section 3 how to implement these ideas in the case of the coupling deformation, which is considerably easier since it is an exact deformation. Another
example, that is expected to work in a similar fashion is the $\beta$ deformation of ${\cal N}=4$ \cite{beta}. More general deformations may also be considered. 
Whether this technique will be useful in unveiling new integrability structures in the perturbative computation of the couplings  $a_{ABC}$
remains an open problem.

At strong 't Hooft coupling we may use the AdS/CFT duality \cite{AdS/CFT} to compute the couplings $a_{ABC}$. The duality relates the $AdS$ string partition
function, computed  with suitable boundary condition, to the generating functional for correlation functions of the gauge theory \cite{GKP,Witten}. 
However, in practice this relation has been useful only to compute, at strong coupling, correlation functions of chiral operators that are dual to the 
supergravity fields \cite{FMMR,LMRS,AF,Lee:1999pj}. Recently, a path integral approach to compute the
string theory partition function for a heavy string state propagating between two boundary points has been developed \cite{Janik:2010gc} (see also \cite{Buchbinder:2010vw}).
In this case the string  path integral is dominated by a classical  saddle point, giving a
new method to compute at strong coupling the 2-point function of single trace operators with a large number of basic fields \cite{Gubser:2002tv,Frolov:2003qc,Tseytlin:2003ii}.
In Section 4 we shall extend this computation to the case of a 3-point function with an additional chiral operator. 
The basic idea is that, taking into account the coupling between the heavy string worldsheet and the supergravity fields, the 
path integral giving the aforementioned  2-point function can be extended to include the interaction with light fields\footnote{The same idea is explored independently by Zarembo \cite{Zarembo}.}. In practice all one needs to do
is to compute a Witten diagram with a supergravity field propagating from the $AdS$ boundary to the heavy string worldsheet, which acts as a tadpole for this field. 
We will show how this computation works for the dilaton field and several heavy string configurations, obtaining
couplings of the form  $a_{{\cal L}AA}$, in complete agreement with the value
predicted by renormalisation group arguments.

We conclude in Section 5 with comments and open problems.


\section{Three point couplings from linear deformations}

The goal of this section is to show how to relate the 3-point correlation function in a  CFT to the anomalous dimension matrix obtained from deforming the CFT
with a marginal or irrelevant operator ${\cal D}$ of dimension $\Delta$ at the CFT fixed point. 
We emphasise that the results presented in this section are valid at a CFT fixed point with coupling $\lambda$. 
We shall explore basic ideas given in \cite{Cardy} (see also appendix in \cite{Freedman:2005wx}).
The example that we have in mind, and that we will work in detail in the following
sections, is ${\cal N}=4$ SYM, so we shall stick to four dimensions. In this case, the dimension $\Delta$ of the 
operator ${\cal D}$ satisfies $\Delta \ge 4$. 
In the case of ${\cal N}=4$ SYM we have a line of CFT's parameterised by the coupling constant $\lambda$, so we may wish
to take the coupling to be finite  and large, or to expand to arbitrary order in the coupling constant. We may also wish to consider an operator
 ${\cal D}$ of protected dimension, but that is not necessary.

Our starting point is a CFT with action $S$. We consider the deformed theory with action
\begin{equation}
S_u=S + u\, \Lambda^{4-\Delta} \int d^4y \,{\cal D}(y)\,,
\label{DeformedAction}
\end{equation}
where $u$ is the dimensionless deformation parameter at the cut-off scale  $\Lambda$ of dimension inverse length, and the 
operators that appear in this action are the renormalized operators of the undeformed theory. The beta function for the coupling 
$u$ has the form,
\begin{equation}
\beta_u=\frac{d u}{d\, \ln\Lambda} = (\Delta - 4) u + \cdots \,.
\end{equation}
where $\cdots$  represents  terms quadratic, or of higher powers, in the couplings to all operators around the fixed point. 
For what we are doing it will be sufficient to work to linear order in $u$, so we keep only the
first term in the beta function $\beta_u$ with $\Delta$ computed at the fixed point.
Sending the cut-off to infinity, the coupling $u(\mu)$ at a fixed scale  $\mu$ is constant for $\Delta=4$ (marginal deformation) and 
vanishes for $\Delta>4$ (irrelevant deformation). 
For simplicity we shall consider the operator ${\cal D}$ to be a scalar primary. But this can be generalised to more operators, for 
instance,  ${\cal D}$ could be the energy-momentum tensor, in which case $u$ would be a tensor valued deformation parameter.

For the sake of clarity, we shall consider in what follows the case of an operator ${\cal D}$ with dimension $\Delta=4$ at the fixed point.
Since we are interested in the case of ${\cal N}=4$ SYM at any value of the coupling, this means the operator has protected dimension.
In the appendix we extend our results to the case of irrelevant deformations. We decided to separate the discussion because in the following sections we shall
be working with the marginal case, therefore avoiding the duplication of formulae in the main text.

A final introductory word about notation, we shall  use the label $u$ to denote quantities computed in the deformed CFT with action
given by (\ref{DeformedAction}). Quantities without the label $u$ are computed at the undeformed theory for which $u=0$.

\subsection{Analysis of divergences}

We now analyse the divergences that appear in the deformed theory, in terms of renormalized quantities of the undeformed theory. 
Let ${\cal O}_A$ be any renormalized operator of the undeformed theory.  We shall denote its full dimension (classical + quantum),
at the fixed point, by $\Delta_A$. When
computing the correlation function of this operator with any other operators,   we obtain in the deformed theory to linear order in $u$,
\begin{equation}
\left\langle {\cal O}_A(x) \cdots \right\rangle_u
 = \left\langle {\cal O}_A(x) \cdots \right\rangle -  u  \int d^4y \left\langle {\cal O}_A(x)  \,{\cal D}(y)\cdots \right\rangle \,,
 \label{expansionN-pf}
\end{equation}
where the right hand side of this equation is computed in the undeformed theory.
In general new divergences can appear in equation (\ref{expansionN-pf}), that can be cancelled by renormalizing the operators ${\cal O}_A,\cdots$, and that come from the
behaviour of the correlation function involving ${\cal D}(y)$, when $y$ approaches any of the positions of the other operators. 
The form of this divergences 
is entirely determined by the OPE in the undeformed theory 
of the operator ${\cal D}$ with the operators appearing in the correlation function. 
For the operator ${\cal O}_A(x)$ we have
\begin{equation}
{\cal D}(y)\,{\cal O}_A(x) \sim \sum_B
\frac{a_{{\cal D}AB}\,{\cal O}_B(x)}{ |x-y|^{4+\Delta_A-\Delta_B}}\,,
\label{OPE}
\end{equation}
where the constants $a_{{\cal D}AB}$ are precisely the couplings appearing in the 3-point function $\langle  {\cal D}\,{\cal O}_A{\cal O}_B \rangle $. 
We remark that for now we assume that the complete basis of operators $\{ {\cal O}_A\}$ is diagonal with unit norm, i.e.
\begin{equation}
\langle {\cal O}_A(x)\,{\cal O}_B(0)  \rangle = \frac{\delta_{AB}}{|x|^{2\Delta_A}}\,.
\end{equation}
The physically meaningful couplings $a_{{\cal D}AB}$ are defined with respect to
operators satisfying this normalisation.

Using the OPE expansion (\ref{OPE}), we conclude that the 
divergence in the $y$ integral  of (\ref{expansionN-pf}), arising from the region of integration $y\sim x$, is given by
\begin{equation}
 \int \frac{d^4y}{|x-y|^{4+ \Delta_A-\Delta_B}} 
\approx
2\pi^2
\begin{cases}
\ln \left( \Lambda |x| \right)\,, &\ \ \ \ \ \ \Delta_B=\Delta_A\,,\\ 
\displaystyle\frac{ \Lambda^{ \Delta_A-\Delta_B}}{\Delta_A-\Delta_B}  \,,&\ \ \ \ \ \  \Delta_B<\Delta_A\,.
\end{cases}
\label{divergences}
\end{equation}
Hence, powerlike divergences arise from operators  that enter the OPE of ${\cal O}_A$
and ${\cal D}$, and whose dimensions satisfy $  \Delta_B <\Delta_A$.
By the unitarity bounds this is a finite number of operators, for instance, for scalar operators in four dimensions we must 
have $\Delta_B \ge 1$. Logarithmic divergences appear  from operators in the OPE with $\Delta_A=\Delta_B$.

We are now in position to define renormalized operators ${\cal O}^u_A$ of the deformed theory, expressed in terms of renormalized operators
of the undeformed theory, such that the general correlation function
(\ref{expansionN-pf}) is finite. This is quite simple, because there is a finite number of operators ${\cal O}_B$ entering the OPE (\ref{OPE}) 
and contributing to the divergences in (\ref{divergences}). We define the  renormalized operators
\begin{equation}
{\cal O}^u_A = {\cal O}_A +  u \sum_{\Delta_B=\Delta_A}  2\pi^2 a_{{\cal D}AB}  \left(\ln  \Lambda\right) {\cal O}_B + 
u \sum_{\Delta_B<\Delta_A}  2\pi^2a_{{\cal D}AB}  \,\frac{\Lambda^{ \Delta_A-\Delta_B} }{ \Delta_A-\Delta_B} \,{\cal O}_B\,.
\label{renormalized}
\end{equation}
As usual, we see that operator mixing occurs for $\Delta_B\le\Delta_A$.\footnote{If one
writes the renormalized operator ${\cal O}^u_A$  in terms of bare operators of the undeformed theory, then,
in a theory without dimensional couplings, mixing will only occur between operators of the same classical dimension, i.e. for $\Delta_A^0=\Delta_B^0$. Then,
the last term in (\ref{renormalized}) only concerns operators with different anomalous dimensions, since the power like divergence  becomes logarithmic
when expanding in the coupling $\lambda$.}
With this renormalization scheme, correlation functions 
\begin{equation}
\left\langle {\cal O}^u_A(x) \cdots \right\rangle_u\,,
\end{equation}
computed at the fixed value of the coupling $\lambda$,  and to linear order in $u$ for the theory with action (\ref{DeformedAction}), 
are finite.  

Of particular importance to us will be the case of  2-point functions. For operators  ${\cal O}_A$ and ${\cal O}_B$ 
with the same dimension in the undeformed theory, it is simple to see that
\begin{equation}
\left\langle {\cal O}^u_A(x) {\cal O}^u_B(0) \right\rangle_u = 
\frac{1}{|x|^{2\Delta_A}}\Big(  \delta_{AB} -  u\,2\pi^2 \left( a_{{\cal D}AB} +  a_{{\cal D}BA}\right)  \ln |x|\Big)\,.
\end{equation}
For ${\cal O}_A={\cal O}_B$ this gives
\begin{equation}
\left\langle {\cal O}^u_A(x) {\cal O}^u_A(0) \right\rangle_u = \frac{1}{|x|^{2\left(\Delta_A + u\,2\pi^2 a_{{\cal D}AA} \right)}}\,.
\end{equation}
If there are different operators ${\cal O}_A$ and ${\cal O}_B$ of equal dimension, we see that the effect of turning on the deformation is to induce
operator mixing, since the above 2-point function is no longer diagonal. 
It is also simple to see that the  2-point function for operators  ${\cal O}_A$ and ${\cal O}_B$ of different dimension still vanishes.

\subsection{Deformed anomalous dimension matrix}

We now wish to better understand the basis of renormalized operators introduced in the previous section, by defining a deformed 
anomalous dimension matrix. We will then verify the Callan-Symanzik equation for correlation functions in the deformed theory.

Let us start by defining renormalized operators of the deformed theory using the usual renormalization matrix
\begin{equation}
{\cal O}^u_A =  Z_{AB} (\Lambda,u) \,{\cal O}_B \,,
\label{RMatrix}
\end{equation}
where we omitted the summation in $B$.
From (\ref{renormalized}) we can read the entries of this matrix,
\begin{eqnarray}
&&Z_{AA} = \Lambda^{ u \,2\pi^2 a_{{\cal D}AA}}  \,,\\ 
&&Z_{AB} = u \,2\pi^2 a_{{\cal D}AB}
\begin{cases}
\ln\Lambda\,, &\ \ \ \ \ \ \Delta_B=\Delta_A\,,\\
\displaystyle\frac{ \Lambda^{ \Delta_A-\Delta_B}}{\Delta_A-\Delta_B}  \,,&\ \ \ \ \ \  \Delta_B<\Delta_A\,.
\end{cases}
\label{Z}
\end{eqnarray}
It is now simple to compute the anomalous dimension matrix associated to the deformation, defined by
\begin{equation}
\Gamma_{AB} = Z^{-1}_{AC}\, \frac{d\ \ }{d \ln\Lambda} \,Z_{CB}\,.
\label{Gamma}
\end{equation} 
Its non-vanishing entries are
\beq
\Gamma_{AB} = u \,2\pi^2 a_{{\cal D}AB} \,\Lambda^{\Delta_A-\Delta_B}\,,
\label{DeformedGamma}
\eeq
for $\Delta_B \le \Delta_A$. 
We remark that the anomalous dimension matrix $\Gamma_{AB}$ is defined with respect to renormalized operators of the undeformed 
theory with total dimension given by $\Delta_A$ at the fixed point.
If we order the operators in blocks with descending  value of dimension, the non-diagonal top-right blocks of the anomalous dimension matrix $\Gamma_{AB} $
have zero entries. It is then clear that its eigenvalues $\Gamma_{A}$
are independent of the cut-off $\Lambda$, although the eigenvectors do depend in general on $\Lambda$ (when there is mixing
between operators of different dimension). Thus,  in the diagonal basis we have, as usual,  ${\cal O}^u_A =  \Lambda^{\Gamma_A} \,{\cal O}_A $.

An alternative way of deriving the relation between the anomalous dimension $\Gamma_{AB}$ and the couplings $a_{{\cal D}AB} $
is to verify the Callan-Symanzik equation. This is simpler for a marginal  deformation, and to linear order in $u$, because the beta function $\beta_u$ vanishes (in the appendix we consider the 
case of irrelevant deformations). For the non-renormalized two-point function of the deformed theory, computed using renormalized 
operators of the CFT at the fixed point, the Callan-Symanzik equation has the form 
\begin{equation}
\frac{\partial \  \ }{\partial\ln\Lambda} \left\langle  {\cal O}_A(x) {\cal O}_B(0) \right\rangle_u +
\sum_I \Gamma_{AI}  \langle  {\cal O}_I(x) {\cal O}_B(0) \rangle_u +
\sum_I \Gamma_{BI}  \langle  {\cal O}_A(x) {\cal O}_I(0) \rangle_u =0
\end{equation}
Using (\ref{expansionN-pf}) and the form of the divergences given in (\ref{divergences}) this equation is satisfied provided 
(\ref{DeformedGamma}) holds.

For practical perturbative computations it is useful to relate the couplings  $a_{{\cal D}AB} $ to the anomalous dimension
matrix computed with respect to bare operators of the CFT (not renormalized). Let us denote a basis of such operators by $\{ {\cal O}^b_A\}$.
Now assume that we manage to diagonalize the anomalous dimension matrix of the critical theory, so that in the basis  $\{ {\cal O}^b_A\}$ 
we have  ${\cal O}_A = \Lambda^{\gamma_A}  \,{\cal O}^b_A$,
where $\gamma_{A}$ are the  eigenvalues (for instance, in  ${\cal N}=4$ SYM we can use integrability techniques to do that quite effectively).
In this basis, and denoting by
$\Delta_A^0$ the classical dimension of operators, it is simple to see that the renormalization matrix ${\cal Z}_{AB}$ relating bare operators
to the renormalized operators of the deformed theory in the usual way, ${\cal O}^u_A =  {\cal Z}_{AB} \,{\cal O}^b_B $,
has entries
\begin{eqnarray}
&&{\cal Z}_{AA} = \Lambda^{\gamma_A + u \,2\pi^2 a_{{\cal D}AA} } \,,\\ 
&&{\cal Z}_{AB} = u \,2\pi^2 a_{{\cal D}AB}
\begin{cases}
\Lambda^{\gamma_A} \ln\Lambda\,, &\ \ \ \ \ \ \Delta_B=\Delta_A\,,\\
\displaystyle\frac{ \Lambda^{ \Delta^0_A -\Delta^0_B + \gamma_A }}{\Delta_A-\Delta_B}  \,,&\ \ \ \ \ \  \Delta_B<\Delta_A\,.
\end{cases}
\end{eqnarray}
The corresponding deformed anomalous dimension matrix has entries
\beq
{\cal H}^u_{AB} = \delta_{AB} \gamma_A + 
 u \,2\pi^2 a_{{\cal D}AB} \,\Lambda^{\Delta^0_A-\Delta^0_B}\,.
\label{DeformedH}
\eeq
Note that these are the entries of the matrix ${\cal H}^u$ in the basis $\{ {\cal O}^b_A\}$ that diagonalizes the anomalous dimension matrix of the critical theory.
Again, it is important to realize that the structure of the matrix ${\cal H}^u$, given by (\ref{DeformedH}), implies that its eigenvalues are independent of the cut-off
$\Lambda$, although the eigenvectors may depend on $\Lambda$.

Let us show explicitly how the knowledge of the deformed anomalous dimension ${\cal H}^u$ allows to 
relate the couplings $a_{{\cal D}AB}$ to the deformed anomalous dimensions and renormalized operators ${\cal O}^u_A$
expressed in terms of the bare quantities. First we write the anomalous dimension matrix as
\begin{equation}
{\cal H}^u = {\cal H} + u {\cal H}'\,,
\end{equation}
where ${\cal H}$ is the anomalous dimension matrix of the critical theory, and $u {\cal H}'$ is the term arising from the deformation which we treat as a perturbation.
For simplicity we assume that the operators ${\cal O}_A = \Lambda^{\gamma_A}  \,{\cal O}^b_A$ and 
${\cal O}_B = \Lambda^{\gamma_B}  \,{\cal O}^b_B$ do not have the same anomalous dimension 
at the critical point (they may or may not have the same classical dimension).
Then, writing the eigenvalues and eigenvectors of ${\cal H}^u$ respectively as 
\begin{equation}
\gamma^u_A = \gamma_A + u \gamma'_A\,,\ \ \ \ \ \ \ \ 
{\cal O}^u_A = {\cal O}_A + u\, {\cal O}'_A\,,
\end{equation}
basic quantum mechanics formulae gives
\begin{equation} 
\gamma'_A =  \langle  {\cal O}_A | {\cal H}' | {\cal O}_A\rangle \,,\ \ \ \ \ \ \ \ \ 
 {\cal O}'_A= \sum_{A\ne B} \, \frac{\langle {\cal O}_B| {\cal H}' |  {\cal O}_A\rangle}{\gamma_A-\gamma_B}\,{\cal O}_B \,,
\label{QuantumMechanics}
\end{equation}
where the matrix elements are computed in the basis $\{ {\cal O}_A\}$ with unit normalised operators.
From the explicit form of the deformed anomalous dimension matrix  in the basis $\{ {\cal O}^b_A\}$ given in 
(\ref{DeformedH}), we conclude that
\begin{eqnarray} 
&2\pi^2 a_{{\cal D}AA} =  \langle  {\cal O}^b_A | {\cal H}' | {\cal O}^b_A\rangle = \langle  {\cal O}_A | {\cal H}' | {\cal O}_A\rangle  \,, 
\label{final}\\
&2\pi^2 a_{{\cal D}AB} \Lambda^{ \Delta^0_A -\Delta^0_B}=   \langle {\cal O}^b_B| {\cal H}' |  {\cal O}^b_A\rangle=  \Lambda^{\gamma_B-\gamma_A}  \langle {\cal O}_B| {\cal H}' |  {\cal O}_A\rangle\,.
\label{final2}
\end{eqnarray}
Note that (\ref{final2}) has the correct dependence in the cut-off $\Lambda$ to  relate operators ${\cal O}_A$ and ${\cal O}_B $ of different dimension,
as required by (\ref{QuantumMechanics}).

It is now clear that if we have a way of determining the action of the perturbation matrix ${\cal H}'$ on the bare operators, we may then compute the corresponding 
couplings using (\ref{final}) and  (\ref{final2}).  This will be the case in the next section, where we consider coupling deformations of ${\cal N}=4$ SYM and the known form of the integrable anomalous dimension matrix at a given order in the coupling constant.

We finish this section with a word on normalization of operators. In the next section it will actually be convenient to perform computations with operators that are not 
normalized to unit, i.e. after diagonalizing the eigenvectors of the undeformed theory we will have
\begin{equation}
\langle {\cal O}_A(x)\,{\cal O}_B(0)  \rangle = |C_A|^2\,\frac{\delta_{AB}}{|x|^{2\Delta_A}}\,.
\end{equation}
With this normalization, to obtain the physically meaningful couplings for the unit normalized operators, we need to divide the operators in
equations (\ref{final}) and  (\ref{final2}) by their norm.

\section{${\cal N}=4$ SYM and integrability}

In this section we consider the simplest case of ${\cal N}=4$ SYM deformed by the Lagrangian operator, 
since  this theory is actually a line of fixed points parametrised by the coupling constant.
We shall use integrability  to show how to compute the couplings in the 3-point function of the  Lagrangian with
any two operators of the theory.  For simplicity, we restrict our analysis
to the $SU(2)$ scalar subsector, and consider in detail operators corresponding to two-magnon excitations in the spin chain language.

We shall use the following convention for the  ${\cal N}=4$ SYM action
\begin{equation}
S_{{\cal N}=4}=\frac{1}{g_{\mathrm{YM}}^{2}}\int 
d^{4}{y} \,\mathrm{Tr}
\Big(  -\frac {1}{2}\,F_{\mu\nu}F^{\mu\nu} -  D_\mu\phi_I D^\mu\phi^I + \frac{1}{2}\left[ \phi_I,\phi_J\right]^2 + {\rm fermions}\Big)\,,
\label{action}
\end{equation}
where $I=1,\cdots,6$ and the covariant derivative is defined by $D_\mu=\partial_\mu - i [A_\mu,\ ]$.
All fields are in the adjoint representation and the $SU(N)$ generators are normalized with $\mathrm{Tr} \,T^aT^b = \delta^{ab}/2$.
We will be considering the $SU(2)$ sector with complex scalars
\[
Z=\frac{1}{\sqrt{2}}\,\left( \phi_1 + i\phi_2 \right)\,,\ \ \ \ \ \ \ \ \ \ 
X=\frac{1}{\sqrt{2}}\,\left( \phi_3 + i\phi_4 \right)\,.
\] 
Now consider the theory at some fixed value of the 't Hooft coupling, defined by
\begin{equation}
g^2 = \frac{g^2_{YM}N}{16\pi^2} 
= \frac{\lambda}{16\pi^2}
\,.
\end{equation}
We will consider (planar) perturbation theory to some order in the coupling $g^2$. We are therefore considering the 
CFT at the fixed point with coupling $g^2$.  Then, to deform the theory with  ${\cal D}(z) =  {\cal L}(z)$, it is clear from (\ref{action}) that 
we should write
\begin{equation}
g^2\rightarrow g^2 (1 - u) \,.
\end{equation}
Hence, making this replacement in the anomalous dimension matrix of ${\cal N}=4$, to a given order in $g^2$, and then 
keeping only the linear terms in $u$, we obtain the form of the deformed anomalous dimension matrix ${\cal H}^u$.  We may then use the
results (\ref{final}) and (\ref{final2}) to compute the couplings. Alternatively, we can also compute the derivative of $\gamma^u_A$ or ${\cal O}^u_A$ with respect to 
$u$ or, instead, the derivative
\begin{equation} \label{deriv-rule}
\frac{\partial\ }{\partial u} = - g^2 \frac{\partial\ }{\partial g^2}\,
\end{equation}
of  $\gamma_A$ or ${\cal O}_A$  to a given order in $g^2$. Finally, note that the two-point function of the Lagrangian is (dropping the $1/N$ correction)
\begin{equation}
\langle {\cal L}(x){\cal L}(0)\rangle =  \frac{3 N^2}{\pi^4} \,\frac{1}{x^8}\,.
\label{normalisationL}
\end{equation}
We shall compute the couplings with respect to this normalization, but it is simple to  re-scale with respect to the unit normalised Lagrangean $\hat{\cal L}$.
In that case we would obtain that all the
couplings  $a_{\hat{\cal L}AB}$ computed in this paper are of order $1/N$, for fixed 't Hooft coupling, as expected.

As an example 
consider single trace operators made by $L$ fields of the $SU(2)$ sector
and regard the fields $X$ as impurities in the vacuum state ${\cal O}={\rm Tr}Z^L$. For operators  with $M$ impurities, we use 
the integers $x_1,\cdots ,x_M$ to indicate the position of the impurities in the corresponding spin chain, 
\begin{equation}
 |x_1,\cdots,x_M\rangle \equiv  |Z\cdots Z X Z \cdots Z X Z \cdots \rangle \,.
\end{equation}
The anomalous dimension matrix is that of an integrable spin chain and
may be diagonalized by solving the Bethe equations \cite{Minahan:2002ve}. Then
the operator ${\cal O}_A$, with anomalous dimension $\gamma_A$, is given by 
\begin{equation}
{\cal O}_A = \sum_x \psi_{p_1,\cdots,p_M}(x_1,\cdots,x_M)\,|x_1,\cdots,x_M\rangle\,,
\end{equation}
where the wave function $\psi$ is 
parameterized  by the momenta $p_j$ of the magnons, which  in general can be complex, sum to zero mod $2\pi$,
and depend on the 't Hooft coupling.
Then, the contribution of the $j$-th magnon to the anomalous dimension of the operator ${\cal O}_A$ is given by \cite{Berenstein:2002jq,Gross:2002su,Santambrogio:2002sb,Beisert:2004hm,Beisert:2005tm}
\begin{equation} 
\gamma_j(g^2) = \sqrt{1+ 16 g^2 \sin^2\frac{p_j}{2}}-1\,.
\label{GMall-loop}
\end{equation}
This formula is believed to be correct to all orders in the 't Hooft coupling, provided 
wrapping effects, that become important at order $g^{2L}$, are neglected. 

The interactions between magnons are responsible for the dependence of their momenta on the 't Hooft
coupling. This effect appears in the computation of the anomalous dimension at two-loop
order, while is appears at one-loop in the computation of the coupling $a_{{\cal L}AA} $. 
Thus, neglecting wrapping effects,    (\ref{final}) gives  the all-loop result
\begin{equation}
2\pi^2 a_{{\cal L}AA} =- g^2 \frac{\partial\ }{\partial g^2}\sum_{j=1}^M  \gamma_j (g^2)=
- 8g^2\,
\sum_{j=1}^M\, \frac{\sin^2 \frac{p_j}{2}+g^2 \,p_j'\,\sin \frac{p_j}{2}\,\cos \frac{p_j}{2}}{\sqrt{1+16 g^2\sin^2\frac{p_j}{2}}}\,,
\label{AllLoopLAA}
\end{equation}
where prime denotes derivative with respect to $g^2$.
We remark that the normalised coupling $a_{\hat{\cal L}AA}$ scales with $1/N$ as expected. We shall compute in the next section this 
coupling up to order $g^4$, in the simple case of operators with two-magnons.

Next we consider the dilute limit of $L\gg M$. In this limit the magnons propagate freely on the spin chain and their momentum
is trivially quantised as
\begin{equation}
p_j = \frac{2\pi n_j}{L}\,.
\end{equation}
In this case the second term in the numerator of 
(\ref{AllLoopLAA}) can be dropped.
One may now study both weak and strong coupling limits.
The leading order term in $g^2$, which comes from the 1-loop correction to the anomalous dimension of 
${\cal O}_A$, is given by
\begin{equation}
2\pi^2  a_{{\cal L}AA}  \approx
- 8g^2 
\sum_{j=1}^M  \sin^2 \frac{p_j}{2} + O(g^4)\,,
\end{equation}
and can be derived simply by doing Wick contractions
between ${\cal L}$, ${\cal O}_A$ and 
$\bar{\cal O}_A$. On the other hand, at strong coupling and neglecting wrapping effects, we have
\begin{equation} \label{SCpred}
2\pi^2   a_{{\cal L}AA} \approx
-2 g
\sum_{j=1}^M 
\left|\sin \frac{p_j}{2}\right| + O(1)\,.
\end{equation}
In Section~\ref{sec:strongcoup} we shall confirm this  computation of the coupling $a_{{\cal L}AA} $,
by directly computing this 3-point function 
using the AdS/CFT duality in the gravity limit.

\subsection{Two-magnon operators}

We shall now illustrate how one can use integrability techniques and the general results given in (\ref{final}) and (\ref{final2}) to compute the couplings of two operators, each
with two magnons, and the Lagrangian. We will
compute 1-loop corrections to these couplings, which correspond to diagonalizing the anomalous dimension matrix at two-loop
order. The corresponding spin chain Hamiltonian includes next to neighbour interactions \cite{Beisert:2003tq},
\begin{equation}
{\cal H}=2 g^2\left(1-4 g^2 \right) \sum_{x=1}^L \left(  1 - P_{x,x+1}\right) + 2g^4 \sum_{x=1}^L \left(  1 - P_{x,x+2}\right) \,,
\label{AnomalousDimMatrix}
\end{equation}
where $P$ is the permutation operator. At this order, the Bethe wave function includes a contact term and can be written in the following form \cite{Staudacher:2004tk}
\begin{equation}
\psi_{p_1,p_2}(x_1,x_2) = \phi_{p_1,p_2} (x_1,x_2) + S(p_2,p_1) \,\phi_{p_2,p_1} (x_1,x_2)\,,
\label{WaveFunction}
\end{equation}
with
\begin{equation}
\phi_{p_1,p_2} (x_1,x_2) = e^{ ip_1x_1+ip_2x_2}  \Big( 1 + f(p_1,p_2)\delta_{x_1+1,x_2} \Big)\,.
\end{equation}
Since for two magnons the  total momentum vanishes, we have $p=p_1=-p_2$. 
We shall now write the formulae for the contact function and for the S-matrix
in this simpler case \cite{Staudacher:2004tk}. For the contact function $f(p_1,p_2)  = f(p)$, we have
\begin{equation}
 f(p) = 4 \sin^2\frac{p}{2}\,,
\end{equation}
which satisfies $f(p_2,p_1)  = f(-p)=f(p)$.
The S-matrix  $S(p_1,p_2)=S(p)$, can be written as 
\begin{equation}
S(p) = S^{(0)}(p) + g^2 S^{(1)}(p) \,,
\end{equation}
with
\begin{equation}
S^{(0)}(p) = - \frac{1-  e^{ip} }{1 - e^{-ip} }\,,\ \ \ \ \ \ \ 
S^{(1)}(p) =  8 i \sin^2\frac{p}{2} \sin{p} \,S^{(0)}(p) \,.
\end{equation}
It is clear that $S(p_2,p_1) = S(-p) = 1/S(p)$.

The momenta that solve the Bethe equation $e^{ipL}=S(p)$ are given by
\begin{equation}
p_n = \frac{2\pi n}{L-1} - \frac{16 g^2}{L-1}\,\cos\frac{\pi n}{L-1}\,\sin^3\frac{\pi n }{L-1}\,,
\end{equation}
where $n$ is an integer. It is now a mechanical calculation to replace this expression in (\ref{AllLoopLAA}), to obtain
\begin{equation}
2\pi^2a_{{\cal L}AA} =
-16g^2
 \left[ \sin^2 \frac{\pi n}{L-1} - 8g^2\sin^4\frac{\pi n}{L-1} \left(  1- \frac{4}{L-1}\,\cos^2\frac{\pi n}{L-1} \right)\right] + O(g^6)\,.
\label{TwoLoopLAA}
\end{equation}

Next we consider the coupling $a_{{\cal L}AB} $, where  ${\cal O}_A$ is an operator with two magnons of momenta $p$ and $-p$, and 
${\cal O}_B$ is an operator with two magnons of momenta $q$ and $-q$. This amounts to computing
the matrix element $\langle {\cal O}^b_B| {\cal H}' |{\cal O}^b_A\rangle$. Using the two-loop anomalous dimension matrix 
given in (\ref{AnomalousDimMatrix}) we have
\begin{equation}
{\cal H}'= - {\cal H} + 8g^4\sum_{x=1}^L \left(  1 - P_{x,x+1}\right) - 2g^4 \sum_{x=1}^L \left(  1 - P_{x,x+2}\right) \,.
\label{DerivativeH}
\end{equation}
Now we argue that the first two terms in this expression do not contribute to $\langle {\cal O}^b_B| {\cal H}' |{\cal O}^b_A\rangle$.
First recall that $|{\cal O}^b_A\rangle$ and $|{\cal O}^b_B\rangle$ are eigenstates of ${\cal H}$, with terms of order  $g^0$ and  $g^2$. 
Since ${\cal H}$ is diagonalized by these eigenstates, the  contribution from the first term in ${\cal H}'$ vanishes.
Moreover, since
the second term in ${\cal H}'$ starts at order $g^4$, for this term we may consider the eigenstates $|{\cal O}^b_A\rangle$ and $|{\cal O}^b_B\rangle$  only at order $g^0$. 
Thus, this term is proportional to the  Hamiltonian ${\cal H}$ at one-loop, and it will also give a vanishing contribution. We are therefore left with the contribution from the last term in 
(\ref{DerivativeH}), which we can compute with the eigenstates $|{\cal O}^b_A\rangle$ and $|{\cal O}^b_B\rangle$  of order $g^0$. 
A computation shows that
\begin{equation}
\langle {\cal O}^b_B| {\cal H}' |{\cal O}^b_A \rangle = 64 g^4 L \,e^{-\frac{i}{2}\left( p-q\right)}\sin\frac{p}{2}\, \sin p\,  \sin\frac{q}{2} \,\sin q  \,.
\end{equation}
Since we are working with states with  norm  $|C_A|^2=L(L-1)+O(g^2)$,  after normalising to unit we obtain 
\begin{equation}
2\pi^2 a_{{\cal L}AB} =64 g^4\, \frac{e^{\frac{i}{2} \left( q-p\right)}}{L-1}\, \sin\frac{p}{2}\, \sin p\,  \sin\frac{q}{2} \,\sin q  \,.
\label{AllLoopLMN}
\end{equation}

The example given in this section shows that one can use integrability of ${\cal N}=4$ SYM to compute quite effectively the couplings of operators to
the Lagrangian. Of course one can try to compute these couplings to higher orders in the 't Hooft coupling $g^2$, to consider operators with more magnons and also 
operators outside the $SU(2)$ sector. Another generalization would be to consider the beta deformation of ${\cal N}=4$ SYM, computing the couplings involving
the operator that generates such deformation. 

It would be very interesting to study the deformed anomalous dimension matrix associated to operators that are not exact, and whose deformation does not lead to an integrable theory. 
In particular, having a representation of ${\cal H}'$ acting on the spin chain associated to  operators of the CFT at the fixed point would allow for  quite effective computations of the corresponding couplings.

\section{Strong Coupling \label{sec:strongcoup}}

\begin{figure}
\centerline{\includegraphics[height=5cm]{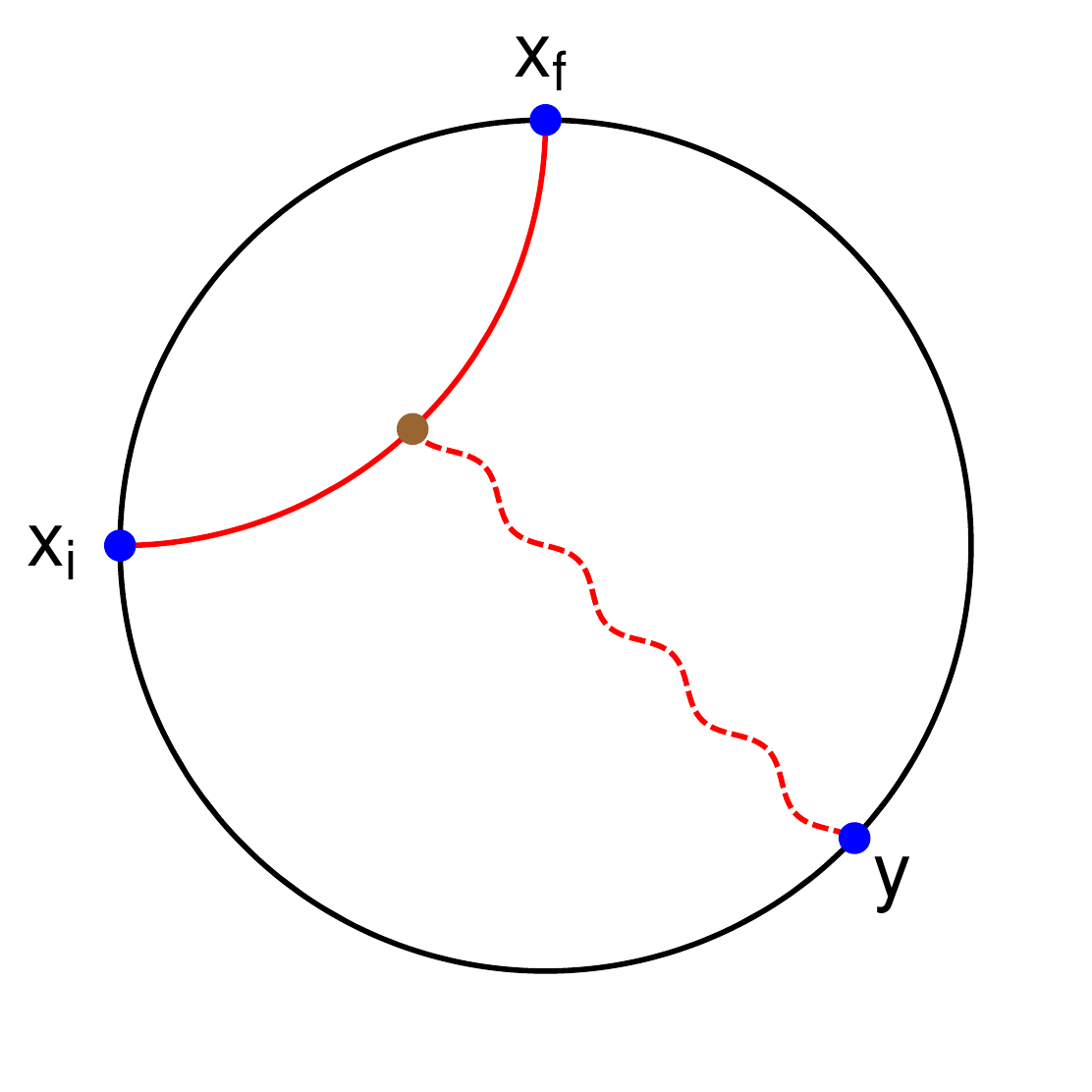}}
\caption{Witten diagram for a 3-point function that represents a heavy string state interacting with a light supergravity field. Note that the heavy string line actually spans
a two-dimensional worldsheet, whose classical saddle point determines the behaviour of the partition function, as explained in \cite{Janik:2010gc}. To leading
order, the string worldsheet acts as a tadpole for the supergravity fluctuations.}
\label{wittendiagram}
\end{figure}

In this section we compute 3-point correlation functions of ${\cal N}=4$ SYM at strong coupling using the AdS/CFT duality. 
So far, computations of correlation functions in the gauge/gravity duality use the field theory limit of strings propagating
in AdS. In this case, the computation of Witten diagrams involves only supergravity fields, giving correlation functions of  chiral operators  \cite{FMMR,LMRS,AF}.
On the other hand, here we shall compute 3-point  correlation functions involving two insertions of an operator ${\cal O}_A$ dual to a very massive string \cite{Gubser:2002tv}, 
with a chiral operator ${\cal D}$ dual to a supergravity field. The corresponding Witten diagram is given in Figure \ref{wittendiagram}, where a heavy string
state propagates between boundary points at $x_i$ and $x_f$, and interacts with a light field sourced at the boundary point $y$.
This computation can be done for the supergravity fields that couple to a heavy string worldsheet.
Clearly one can also generalise this computation to higher point functions with more supergravity fields.

To compute the string partition function we shall use different approaches to treat the heavy and light string fields. 
For the heavy string state we shall consider the action for a string (or particle) in the first quantised theory and
compute its contribution to the partition function by summing 
over classical  trajectories, while for the light fields we shall use the supergravity approximation\footnote{For a related discussion see \cite{Tseytlin:1990vf}.}. 
It is therefore convenient to represent the source for the operator ${\cal O}_A$  dual to the
heavy string field by $J(x)$, and the source for the chiral operators  ${\cal D}$ dual to the supergravity fields by $\Phi_0(y)$. 
The gauge theory generating functional for diagrams with insertions of ${\cal O}_A$ at  $x_i$ and at $x_f$ can then be written as
\begin{equation}
\tilde{Z}(x_i,x_f,\Phi_0) \equiv \left. \frac{\delta^2 Z(J,\Phi_0)}{\delta J(x_i) \,\delta J(x_f)} \right|_{J=0}
=  \left\langle{\cal O}_A(x_i) \,{\cal O}_A(x_f) \,e^{  \int d^4 y \,\Phi_0(y)\, {\cal D}(y)}\right\rangle_{{\cal N}=4}\,.
\label{GeneratingFunctional}
\end{equation}
By varying with respect to the sources $\Phi_0(y)$ we may compute 
correlation functions  with many chiral operators. In this section we are interested in the
simplest case of taking one such derivative to compute the  3-point function $\langle  {\cal O}_A{\cal O}_A{\cal D}\rangle$ for
some chiral operator ${\cal D}$.

The AdS/CFT duality states that the gauge theory generating functional for correlation functions of local operators
equals the string partition function with suitable boundary conditions \cite{GKP,Witten}.
In particular, at strong coupling, the generating functional (\ref{GeneratingFunctional})  can be approximated by
\begin{equation}
\tilde{Z}(x_i,x_f,\Phi_0)
\approx \int DX\, D\gamma\,D\Phi \,e^{i \left(S_{P}[X,\gamma,\Phi] +  S_{SUGRA}[\Phi] \right)}\,,
\label{StringGeneratingFunctional}
\end{equation}
where we use the string Polyakov action $S_P$ to describe the propagation of the heavy string state.
The corresponding  worldsheet starts and ends very close to the boundary, i.e. 
 in Poincar\'e coordinates $x^a=(x^\mu,z)$ it obeys the boundary conditions
\begin{equation}
X^a(\tau_i,\sigma) = x^a_i=\left(x_i^\mu,\epsilon\right)\,,\ \ \ \ \ \ \ \ \ 
X^a(\tau_f,\sigma) = x^a_f=\left(x_f^\mu,\epsilon\right)\,,
\label{bconditions}
\end{equation}
where $\epsilon$ is a regulator. The effect of these boundary conditions 
is to generate two functional derivatives with respect to the source $J(x)$ of the heavy field,
justifying the identification between (\ref{GeneratingFunctional}) and (\ref{StringGeneratingFunctional}).
The supergravity fields in (\ref{StringGeneratingFunctional}) are represented by $\Phi$ and
approximate the gauge theory sources $\Phi_0(x^\mu)$ near the boundary, in the sense that
$\Phi(x^\mu,\epsilon) \rightarrow  \epsilon^{4-\Delta} \Phi_0(x^\mu)$ as $\epsilon\rightarrow 0$. 
 
The propagation of the light fields is determined by the supergravity action around the 
$AdS_5\times S^5$ vacuum, which we denote below by $\Phi=0$.  The vacuum
value for the ten-dimensional Einstein metric $g_{AB}$ is given by 
\begin{equation}
ds^2 = \ell^2g_{AB} \, dx^A dx^B = \ell^2\,\frac{dx^\mu dx_\mu + dz^2}{z^2} + \ell^2\,d\Omega_5^2\,,
\end{equation}
where the AdS radius satisfies $\ell^4= 4\pi g_s N \alpha'^2$. Then, it is simple to show that
the five-dimensional supergravity action in the Einstein's frame has the form
\begin{equation}
S_{SUGRA} = \frac{1}{2 \kappa^2} \int d^5x \sqrt{-g} \left( R +12 -\frac{1}{2}\,(\partial \phi)^2 + \cdots \right)\,,
\end{equation}
where $\cdots$ represents  terms in the action other than the  metric  and dilaton fields. The gravitational
coupling is given by  $\kappa = 2\pi/N$.

The propagation of the heavy string state, and its coupling to the supergravity fields, is determined from the Polyakov
action
\beq
S_P[X,\gamma,\Phi]  = - g \int d^2 \sigma \, \sqrt{-\gamma}\, \gamma^{\alpha \beta} \, \partial_\alpha X^A \partial_\beta X^B \, g_{AB} \, e^{\phi/2}+\cdots \,,
\label{Polyakov}
\eeq
where $g_{AB}$ is the ten-dimensional metric in the Einstein frame, $\phi$ represents  the  fluctuations of the dilaton field and $\cdots$ includes other terms like 
worldsheet fermions and other supergravity fields. The heavy string will have the worldsheet topology of a cylinder. Working in the conformal gauge, the integration
over worldsheet metrics becomes simply an integration over the modular parameter $s$ of the cylinder, i.e. 
\beq
\label{gammatos}
\int d^2 \sigma \, \sqrt{-\gamma}\,\gamma^{\alpha\beta} \ \ \rightarrow\  \int_{-s/2}^{s/2}d\tau\int_0^{2\pi} d\sigma\,\eta^{\alpha\beta}\,.
\eeq

To compute the generating functional (\ref{StringGeneratingFunctional}) it is convenient to perform first the path integral
over the supergravity fields. We write (\ref{StringGeneratingFunctional}) as
\begin{equation}
\tilde{Z}(x_i,x_f,\Phi_0)
\approx \int DX\, ds\, e^{i S_{P}[X,s,\Phi=0] }
\int D\Phi \,e^{i \left(S_{SUGRA}[\Phi]  + \int d^2\sigma\left. \frac{\delta S_{P}[X,s,\Phi]}{\delta \Phi}\right|_{\Phi=0}\,\Phi + \cdots\right)}  \,.
\label{StringGeneratingFunctional2}
\end{equation}
For a fixed off-shell string worldsheet, the supergravity functional  can be 
computed with Witten diagrams, after boundary sources for the supergravity fields are specified.
The new ingredient are the extra terms localized along the string worldsheet that add to the supergravity action.
These terms determine the coupling between the light fields and the heavy string state. In (\ref{StringGeneratingFunctional2}) we wrote
just the leading term, which acts as a simple tadpole for the supergravity fields (it comes from  a cubic interaction in string field theory).
These terms can be treated  perturbatively and do not affect the free propagators of light fields.

Before computing diagrams with a supergravity field, let us recall the computation of the 2-point function 
for the operator ${\cal O}_A$ dual to a heavy field, as done in \cite{Janik:2010gc}. To obtain the correct 
scaling of the 2-point function it is necessary to convolute the generating functional (\ref{StringGeneratingFunctional2})
with the wave function of the classical field we are considering. In the WKB approximation this amounts to
changing the measure in the string path integral such that the action determining the propagator of the 
heavy field is actually
\begin{equation}
\tilde{S}_P = S_P  - \int_{-s/2}^{s/2} d\tau \int_0^{2\pi} d\sigma 
\left[\,\big(\Pi-\Pi_0\big)^a \big( \dot{X} - \dot{X}_0\big)_a  + \Pi^i\dot{X}_i\right]
\,,
\label{NewAction}
\end{equation}
where we use letters $a$  and $i$ respectively for the $AdS_5$ and $S^5$ indices.
The worldsheet canonical momentum  is $\Pi$, and $\Pi^a_0$ and $ \dot{X}^a_0$ are the $AdS_5$ zero modes
\begin{equation}
\Pi^a_0 = \frac{1}{2\pi} \int_0^{2\pi} d\sigma\,\Pi^a(\tau,\sigma)\,,\ \ \ \ \ \ \ \ 
\dot{X}^a_0 = \frac{1}{2\pi} \int_0^{2\pi} d\sigma \,\dot{X}^a(\tau,\sigma)\,.
\label{ZeroModes}
\end{equation}
The arbitrariness in the definition of these zero modes requires a precise prescription. In \cite{Janik:2010gc} it was proposed to
use the embedding coordinates of $AdS_5$, therefore preserving the $SO(2,4)$ symmetry of the conformal group. For a number of particular
examples, it was shown in \cite{Janik:2010gc} the following result, which is expected to be general,
\begin{equation}
\tilde{Z}(x_i,x_f,\Phi_0=0) \approx
\int DX\, ds\, e^{i \tilde{S}_{P}[X,s,\Phi=0] }\approx  \frac{P}{|x_i-x_f|^{2\Delta_A}}\,,
\end{equation}
where we absorbed the cut-off dependence in the measure. The path integral is dominated by the classical saddle point,
which yields the correct conformal dependence for the 2-point function of the operator ${\cal O}_A$. 
The pre-factor $P$, which is associated to the integration of fluctuations of the 
classical solution, will define the normalization of ${\cal O}_A$. Since the 3-point function is defined 
with respect to unit normalized operators, we shall see below that to leading order we actually do not need to evaluate this pre-factor.

Next let us consider the  3-pt function $\langle  {\cal O}_A{\cal O}_A{\cal D}_\chi\rangle$,
where ${\cal D}_\chi$ is a chiral operator of dimension $\Delta$ dual to some particular supergravity field $\chi$. This field may have some tensor structure in $AdS_5$ and also 
some KK structure from the $S^5$ compactification. 
The functional integral for the supergravity fields in \reef{StringGeneratingFunctional2} can be computed using Witten diagrams. If the field $\chi$ has a source at the boundary, \reef{StringGeneratingFunctional2} leads to
\beq
\label{3pt-function}
\left. \frac{\delta \tilde{Z}(x_i,x_f,\Phi_0)}{\delta \chi_0(y)} \right|_{\Phi_0=0}\approx \int DX\, ds\, e^{i \tilde{S}_{P}[X,s,\Phi=0]}\,I_\chi [X,s;y]    \,,
\eeq
where
\beq
I_\chi [X,s;y] =  i \int_{-s/2}^{s/2} d\tau \int_0^{2\pi} d\sigma  \left. \frac{\delta S_{P}[X,s,\Phi]}{\delta \chi} \right|_{\Phi=0} \, K_\chi(X(\tau,\sigma); y)\,,
\label{I_integral}
\eeq
and $K_\chi(X(\tau,\sigma); y)$ is the bulk-to-boundary propagator of the field $\chi$.
Equation (\ref{3pt-function})  states that 
the  3-point function is simply the expectation value over the heavy string trajectories
of the interaction term $I_\chi [X,s;y]$, weighted by the action $\tilde{S}_P$. Note that 
the measure used for the propagation of the heavy string is that defined by the computation of the 2-point function as described above, i.e. after the convolution with the
heavy state wave function. On the other hand, the coupling $I_\chi [X,s;y]$ is determined by the Polyakov action $S_P$.
As usual, to compute this path integral one expands around the classical saddle point
\beq
X(\tau,\sigma) = \bar{X} (\tau,\sigma) + \frac{\delta X(\tau,\sigma) }{\sqrt{g}}\,, \ \ \ \ \ \ \ \ \ \ \ s=\bar{s} + \delta s\,,
\eeq 
where we rescaled the quantum fluctuations for $X(\tau,\sigma) $ so that the 't Hooft coupling $g^2$ does not enter in the quadratic terms arising from the expansion of the action $\tilde{S}_P$
around the saddle point solution $\bar{X}(\tau,\sigma)$. It is then clear that, after expanding the   interaction term $I_\chi [X,s;y]$ around this saddle point, the dominant contribution
in  (\ref{3pt-function}) for large $g$ is given by
\beq
\left. \frac{\delta \tilde{Z}(x_i,x_f,\Phi_0)}{\delta \chi_0(y)} \right|_{\Phi_0=0} \approx \frac{P}{|x_i-x_f|^{2\Delta_A}} \, I_\chi [\bar{X},\bar{s};y]    \,,
\eeq
where the pre-factor $P$ coincides precisely with that in the computation of the 2-point function of ${\cal O}_A$. Thus, defining  ${\cal O}_A$
to have a unit normalised 2-point function, we conclude that at strong coupling
\beq
\label{result}
\left\langle {\cal O}_A(x_i){\cal O}_A(x_f){\cal D}_\chi(y) \right\rangle \approx \frac{I_\chi [\bar{X},\bar{s};y] }{|x_i-x_f|^{2\Delta_A}}   \,.
\eeq
Equation (\ref{result}) is one of the main results of this paper. The approximations that led to (\ref{result}) assume that the initial and final
heavy string states are the same. This means that interactions with supergravity fields that change conserved charges of the
heavy string, such as $R$-charge or $AdS$ spin, are not taken into account. It would be interesting to consider a heavy string with different
initial and final boundary conditions and include the effect of the light supergravity field on the string saddle point.

To fix our conventions let us remark that  in the simple case of a scalar field $\chi$,
normalised such that  
\begin{equation}
S_{\chi} = - \frac{1}{2} \int d^5x \sqrt{-g} \,\Big( (\partial \chi)^2 +\Delta(\Delta-4)\chi^2 \Big)\,,
\end{equation}
its bulk-to-boundary propagator is given by 
\begin{equation}
\label{Kprop}
 K_\chi(x^\mu,z; y^\nu) = \frac{\Gamma(\Delta)}{\pi^2\Gamma(\Delta -2 )}\,\left( \frac{z}{z^2+ (x-y)^2}\right)^\Delta\,.
\end{equation}
In this simple case the normalisation of the 2-point function of the operator ${\cal D}_\chi$ appearing in (\ref{result}) is given by
\beq
\langle {\cal D}_\chi(x){\cal D}_\chi(y) \rangle = \frac{\Gamma(\Delta+1)}{\pi^2\,\Gamma(\Delta-2)}\,\frac{1}{|x-y|^{2\Delta}} \,.
\label{2pt-functionD}
\eeq

In the remainder of this section we shall compute 3-point functions of the type $\langle {\cal O}_A{\cal O}_A{\cal L}\rangle$. We will consider the simplest
case where the operator ${\cal D}$ is dual to the dilaton field, i.e. we will consider the operator ${\cal D}_\phi={\cal L}$.
This will allow us to check our results since, as shown in Sections 2 and 3, this correlation function
can be obtained from the derivative of  $\langle {\cal O}_A{\cal O}_A\rangle$ with respect to the coupling constant.
We need to be careful with normalisations, since the dilaton field in the SUGRA action has a factor of $\eta=1/(2\kappa^2)$ multiplying the canonical kinetic
term. Instead, we should compute the Witten diagram  with the
canonically normalised field $\tilde{\phi}=\sqrt{\eta} \,\phi$, whose propagator is given by (\ref{Kprop}) with $\Delta=4$.
The final result should then be multiplied by $\sqrt{\eta} $, since ${\cal D}_\phi = \sqrt{\eta}\, {\cal D}_{\tilde{\phi}}$. In practice, 
when computing $I_{\phi}$ in  (\ref{I_integral}), 
this amounts to
taking the derivative of the action $S_P$  with respect to $\phi$, while using the normalised propagator
$K_{\tilde{\phi}}$ as given in (\ref{Kprop}). 
In what follows we shall refer to $I_{\phi}$ in  (\ref{I_integral}) with that abuse of notation.
Finally, let us remark that in our conventions the 
2-point function of ${\cal L}$ is given at large $N$ by (\ref{normalisationL}), which can also be verified at strong coupling using the duality.

\subsection{Point-like string}

Let us consider first the limit where the heavy string field dual to the operator ${\cal O}_A$
can be approximated by a point-particle of mass $m$. 
In the Einstein frame, the Nambu-Goto action for a particle coupled to the dilaton takes the form
\begin{equation}
S_{NG} [X,\Phi]= - m  \int_0^1 d\tau \, e^{\phi/4}  \sqrt{ - \dot{X}^A \dot{X}^B g_{AB} }  \,,
\end{equation}
where dot denotes derivative with respect to the 
worldline parameter $\tau$. On dimensional grounds one concludes that massive string states will have $m \sim g^{1/2}$.
We shall be working with the usual Poincar\'e coordinates $X^a=(x^\mu,z)$ and for simplicity assume only motion in the $AdS_5$ part of the space. 
The corresponding Polyakov action, depending on both the particle trajectory and the einbein ${\rm e}$, is
\begin{equation}
\label{Speinbein}
S_{P} [X,{\rm e},\Phi]=  \frac{1}{2} \int_0^1 d\tau  \, e^{\phi/4} \left( \frac{1}{{\rm e}} \,\dot{X}^a \dot{X}^b g_{ab} - {\rm e}\, m^2 \right) \,.
\end{equation}
The functional integration over the einbein ${\rm e}$ can be substituted by a simple integration over the modular parameter $s$,
\begin{equation}
\label{Sps}
S_{P} [X,s,\Phi]=  \frac{1}{2} \int_{-s/2}^{s/2} d\tau  \, e^{\phi/4} \left(\dot{X}^a \dot{X}^b g_{ab} - m^2 \right) \,,
\end{equation}
analogously to \reef{gammatos}. We may now apply the procedure to obtain the 3-point 
function starting from \reef{StringGeneratingFunctional2}.

For spacelike separation on the boundary along a direction $x$, the particle action on the $AdS$ vacuum simplifies to 
\beq
\label{Spnodilaton}
S_P[X,s,\Phi=0] = \half \int_{-s/2}^{s/2} d\tau \left( \frac{\dot{x}^2 +\dot{z}^2}{z^2} - m^2 \right)\,.
\eeq
The computation of the 2-point function for the point particle, using this action, was performed in \cite{Janik:2010gc}. The procedure is
as follows: (i) Determine a solution $\bar{X}$ to the particle equations of motion,
\begin{align}
\label{xzparticle}
x(\tau) &= R\, \tanh \kappa \tau +x_0 \,, \nonumber \\
z(\tau) &= \frac{R}{\cosh \kappa \tau}\,;
\end{align}
(ii) Impose that the endpoints of the motion approach the boundary, $z(\pm s/2)=\epsilon$, which implies
\beq
\label{kappabc}
\kappa \approx \frac{2}{s} \log{\frac{x_f}{\epsilon}}\,,
\eeq
where we have set $x_i = x(-s/2)=0$ and $x_f  = x(s/2) \approx 2\, R \approx 2\, x_0\,$; (iii) Compute  the action
\beq
S_P[\bar{X},s,\Phi=0] = \half \left( \frac{4}{s^2} \log^2\frac{x_f}{\epsilon} - m^2 \right)s\,;
\eeq
(iv) Perform the integration over the modular parameter $s$ by taking the saddle point,
\beq
\label{saddleparticle}
\bar{s} = - i \, \frac{2}{m} \log{\frac{x_f}{\epsilon}}\,,
\eeq
which corresponds to the ``Virasoro constraint'' for the einbein. This computation leads to the correct dependence of the 2-point function,
because at the saddle point
\beq
e^{i S_P[\bar{X},\bar{s},\Phi=0]} = \left( \frac{\epsilon}{x_f} \right)^{2\Delta_A}\,,
\eeq
where we considered the large $\Delta_A$ limit, for which $m \approx \Delta_A$.

To compute the 3-point function $\langle {\cal O}_A {\cal O}_A {\cal L}\rangle$, we need to evaluate $I_\phi [X,s;y] $, as given by (\ref{I_integral}). Taking
care of the correct normalization, we have
\beq
I_\phi [X,s;y] = i\,\frac{3}{4\pi^2} \int_{-s/2}^{s/2} d\tau \left( \frac{\dot{x}^2 +\dot{z}^2}{z^2} - m^2 \right) \left( \frac{z}{z^2+ (x-y)^2}\right)^4\,.
\eeq
For small $\epsilon$, at the saddle point trajectory (\ref{xzparticle}) we obtain 
\beq
I_\phi[\bar{X},s;y] = \frac{i}{32\pi^2}\, \frac{\left( \frac{4}{s^2} \log^2\frac{x_f}{\epsilon} - m^2 \right)s}{\log\frac{x_f}{\epsilon}}\; \frac{x_f^4}{y^4\,(x_f-y)^4 } \,.
\eeq
At the modular parameter saddle point \reef{saddleparticle}, this expression becomes simply
\beq
I_\phi [\bar{X},\bar{s};y] =- \frac{m}{8 \pi^2} \;\frac{x_f^4}{y^4\,(x_f-y)^4 } \,.
\eeq
We conclude from (\ref{result}) that
\beq
\langle {\cal O}_A(0){\cal O}_A(x_f){\cal L}(y) \rangle \approx - \frac{\Delta_A}{8 \pi^2} \;\frac{1}{ x_f^{2\Delta_A-4}\,y^4\,(x_f-y)^4 }\,.
\label{ResultParticle}
\eeq
This expression has the spacetime dependence required by conformal invariance. The coupling $a_{{\cal L}AA}$ is determined for large $\Delta_A$, and it agrees with the expectation from the renormalization group result (\ref{final}). To see this, notice that since $\Delta_A \approx m \sim g^{1/2}$, we have
\beq
2\pi^2  a_{{\cal L}AA} = - 
g^2 \frac{\partial \Delta_A}{\partial g^2} \approx -\frac{\Delta_A}{4} \,.
\eeq
in agreement with (\ref{ResultParticle}).

\subsection{Circular rotating string}

The simplest example after the point particle is the circular rotating string with two equal spins \cite{Frolov:2003qc}, whose 2-point 
function was also computed in \cite{Janik:2010gc}. We start with the Polyakov action coupled to the metric and the dilaton field (\ref{Polyakov}).
The solution $\bar{X}$ for the circular rotating string is given by \reef{xzparticle} in the $AdS_5$ part of the geometry. In the $S^5$ part, with line element
\beq
ds^2_{S^5} = d\gamma^2 + \cos^2\gamma \,d\phi_3^2 + \sin^2\gamma \left(d\psi^2 + \cos^2\psi \,d\phi_1^2+ \sin^2 \psi \,d\phi_2^2 \right)\,,
\eeq
it is given by
\beq
\gamma = \frac{\pi}{2}\,, \quad \phi_3=0\,, \quad \psi=\sigma\,, \quad \phi_1=\phi_2=\omega \tau\,.
\eeq
The conserved angular momenta of the solution are $J\equiv J_1=J_2=  (2\pi g)\,  \omega $. This configuration is dual to an operator of the type 
${\cal O}_A \sim {\rm Tr} \left( X^{J_1}Z^{J_2}\right)$.

Let us apply now the procedure in \cite{Janik:2010gc}. We have
\begin{align}
S_P[\bar{X},s,\Phi=0] & = g \int_{-s/2}^{s/2} d\tau \int_{0}^{2\pi} d\sigma \left( \frac{\dot{x}^2 +\dot{z}^2}{z^2}
- \psi'^2 + \cos^2 \psi \,\dot{\phi_1}^2 + \sin^2 \psi \,\dot{\phi_2}^2 \right) \nonumber \\
& = 2 \pi g \left( \frac{4}{s^2} \log^2\frac{x_f}{\epsilon} +(\omega^2-1) \right)s \,,
\end{align}
where prime denotes derivative with respect to $\sigma$, and we have imposed the relation \reef{kappabc}. As detailed in \cite{Janik:2010gc}, there is a subtlety in obtaining the string propagator, so that the classical solution for the cylinder coincides with the classical state. This amounts to considering \reef{NewAction},
\beq
\tilde{S}_P[\bar{X},s,\Phi=0] = 2 \pi g \left( \frac{4}{s^2} \log^2\frac{x_f}{\epsilon} -(1+\omega^2) \right)s  \,.
\eeq
The saddle point in the modular parameter $s$ is given by
\beq
\label{saddlecircular}
\bar{s} = - i \, \frac{2}{\sqrt{1+\omega^2}} \log{\frac{x_f}{\epsilon}}\,.
\eeq
Looking  at  \reef{kappabc} this implies the Virasoro constraint $\kappa = i\sqrt{1+\omega^2}$. We conclude that 
at the saddle point, we have
\beq
e^{i \tilde{S}_P[\bar{X},\bar{s},\Phi=0]} = \left( \frac{\epsilon}{x_f} \right)^{8\pi g \,\sqrt{1+\omega^2}}\,.
\eeq
This gives the correct dimension $\Delta_A = 4\pi g \sqrt{1+\omega^2}=2\sqrt{(2\pi g)^2 + J^2}$.

Now we will obtain the 3-point function. First we evaluate
\begin{align}
I_\phi[\bar{X},s;y] & = i\, \frac{3\, g}{\pi^2} \int_{-s/2}^{s/2} d\tau \int_{0}^{2\pi} d\sigma \left( \frac{\dot{x}^2 +\dot{z}^2}{z^2}
- \psi'^2 + \cos^2 \psi \,\dot{\phi_1}^2 + \sin^2 \psi \,\dot{\phi_2}^2 \right) \times \nonumber \\
& \hspace{5cm} \times \left( \frac{z}{z^2+ (x-y)^2}\right)^4 \nonumber \\
& = i\,\frac{g}{4\pi}\, \displaystyle{ \frac{\left( \frac{4}{s^2}\log^2\frac{x_f}{\epsilon}+(\omega^2-1)\right)s}{\log\frac{x_f}{\epsilon}}\; \frac{x_f^4}{y^4\,(x_f-y)^4 }}  \,,
\end{align}
At the saddle point \reef{saddlecircular}, we have
\beq
I_\phi[\bar{X},\bar{s};y] =- \frac{g}{\pi\sqrt{1+\omega^2}} \; \frac{x_f^4}{y^4\,(x_f-y)^4 }\,.
\eeq
Therefore, we conclude that
\beq
\langle{\cal O}_A(0) {\cal O}_A(x_f){\cal L}(y) \rangle \approx - \frac{g}{\pi\sqrt{1+\omega^2}} \;
\frac{1}{ x_f^{\,8\pi g\sqrt{1+\omega^2}-4}\,y^4\,(x_f-y)^4 }\,.
\eeq
As happened in the point particle case, the spacetime dependence is the one required by conformal invariance. The coupling $a_{{\cal L}AA}$ determined in this way agrees with the expectation,
since
\beq
2 \pi^2 a_{{\cal L}AA} = - g^2 \frac{\partial \Delta_A}{\partial g^2}\approx- g^2 \frac{\partial \ }{\partial g^2}\, 2\sqrt{(2\pi g)^2 + J^2} =
-\frac{2\pi\,g}{\sqrt{1+\omega^2}} \,,
\eeq
where we kept the angular momentum $J$ fixed when taking the derivative.

\subsection{Giant Magnon}

Following the same steps, we move to the more complicated case of the giant magnon where the string rotates 
on a $R  \times S^2$ subspace of $AdS_5 \times S^5$ \cite{Hofman:2006xt}. We remark that although an operator with a single magnon
is not  gauge invariant, we are implicitly computing the contribution of a single magnon to 
the 3-point coupling $a_{{\cal L}AA} $ involving an operator ${\cal O}_A$ in the dilute limit.
Since the contribution of a magnon to the  2-point function of some operator was not computed in  \cite{Janik:2010gc},
we will present first this calculation and then concentrate on the 3-point function.

Let us start by writing the solution in Poincar\'e coordinates. The $AdS$ part is the same as in the two 
previous cases given in \eqref{xzparticle}. Parametrizing the $S^5$ as
\beq
ds^2_{S^5} = d\theta^2 + \sin^2\theta \,d\varphi^2 + \cos^2\theta\, d\Omega_3^2\,,
\eeq
the giant magnon has non-trivial worldsheet fields in the  $S^2$ part, given by
\begin{eqnarray}
\label{Poincare-GM}
\cos \theta = \sin \frac{p}{2} \text{sech} (\omega u) \, , \qquad
 \tan \left(\varphi - \omega \tau \right) = \tan \frac{p}{2} \tanh (\omega u)\,,
\end{eqnarray}
where
\begin{equation}
\label{defu}
u = \left(\sigma - \tau \cos \frac{p}{2} \right) \csc \frac{p}{2}\, ,
\end{equation}
and $p \in [0,2 \pi)$ is the momentum of the magnon. The Virasoro constraints,
which we will not impose at this stage, require $\kappa=i\,\omega$, where $\kappa$ given in \eqref{xzparticle} characterises the 
$AdS$ motion.  Then we have
\begin{align}
\label{Polyakov-GM}
S_P [\bar{X},s,\Phi=0]
&= g \int_{-s/2}^{s/2} d\tau \int_{-L}^{L} d\sigma \left( \frac{\dot{x}^2 +\dot{z}^2}{z^2}
+ (\dot{\theta}^2 - \theta'^2) + \sin^2 \psi \,(\dot{\varphi}^2 - \varphi'^2) \right) \nonumber \\
&= g \int_{-\frac{s}{2}}^{\frac{s}{2}}d\tau \int_{-L}^{L} d\sigma \left[\kappa^2 + \omega^2 - 2 \omega^2 \cosh^{-2} (\omega u)\right]\, . 
\end{align}
Using \eqref{NewAction} we
convolute with respect to the wave function of the rotating string state, which will 
change the $S^5$ action into its energy.  We obtain
\begin{equation}
\tilde{S}_P[\bar{X},s,\Phi=0]= g \int_{-\frac{s}{2}}^{\frac{s}{2}}d\tau \int_{-L}^{L} d\sigma \left(\kappa^2 - \omega^2 \right)
= 2\, g \, s \left(\kappa^2 - \omega^2 \right) L \, . 
\end{equation}
Taking into account the condition \eqref{kappabc} for $\kappa$, it is possible to perform the remaining 
integration over the modular parameter $s$ by saddle point, with the result
\beq
\label{saddle-GM}
\bar{s} = -i\, \frac{2}{\omega} \log{\frac{x_f}{\epsilon}} \,.
\eeq
Again, this corresponds to the Virasoro constraint, which in this case reads 
$\kappa=i\,\omega$, and leads to
\beq
\label{Conv-Polyakov-GM}
\tilde{S}_P[\bar{X},\bar{s},\Phi=0]= i\, 8\, g\, \omega \, L \log{\frac{x_f}{\epsilon}} \, . 
\eeq
It is convenient now to introduce the angular momentum,
\begin{align} 
J &= 2 \, g \int_{-L}^{L} d\sigma \, \sin^2\theta \,\dot{\varphi} = 2 \, g
\int_{-L}^{L} d\sigma \, \omega \tanh^2 {(\omega u) } \nonumber \\
& = 4g \left( \omega L - \sin{\frac{p}{2}} \;
\frac{ \sinh{\left(2 \omega L \csc{\frac{p}{2}} \right)} }{ \cos{\left(2 \omega\, i\tau \cot{\frac{p}{2}}\right)} + \cosh{\left(2 \omega L \csc{\frac{p}{2}}\right)} } \right) \nonumber \\
& \approx 4g \left( \omega L - \sin{\frac{p}{2}} \right)
\, ,
\label{Jmagnon}
\end{align}
where we took the large $L$ approximation (notice that the above saddle point defines a $\tau$ integration with 
$i\tau$  real). Substituting in \eqref{Conv-Polyakov-GM} and exponentiating, we obtain the expected behaviour 
for the 2-point function,
\beq \label{2p-GM}
e^{i \tilde{S}_P[\bar{X},\bar{s},\Phi=0]} = \left( \frac{\epsilon}{x_f} \right)^{2 \left( J +4 g\sin \frac{p}{2} \right)}\,,
\eeq
in particular, $\Delta_A = J+4 g\sin \frac{p}{2} $, which agrees with \reef{GMall-loop}.

Now it is straightforward to compute the 3-point function. We evaluate
\begin{align} 
I_\phi[\bar{X},\bar{s};y] &= i\, \frac{3\, g}{\pi^2} \int_{-\bar{s}/2}^{\bar{s}/2} d\tau \int_{-L}^{L} d\sigma 
\Big( \kappa^2 + \omega^2 - 2 \omega^2 \cosh^{-2} (\omega u) \Big) 
\left( \frac{z}{z^2+ (x-y)^2}\right)^4  \nonumber \\
&= \frac{12\,g}{\pi^2} \,\sin \frac{p}{2}\,
\int_{-\bar{s}/2}^{\bar{s}/2} d\tau \; \frac{ \sinh{\left(2 \omega L \csc{\frac{p}{2}} \right)} }{ [\cos{\left(2 \omega\, i\tau \cot{\frac{p}{2}}\right)} + \cosh{\left(2 \omega L \csc{\frac{p}{2}}\right)}]} \times\\
& \phantom{aaaaaaaaaaaaa} \times \frac{x_f^{4}}{ \left[(2y^2-2y\, x_f +x_f^2) \cosh(\omega\, i\tau)+(x_f^2-2y x_f)\sinh(\omega\,i\tau)\right]^4}\,. \nonumber 
\end{align}
Taking the large $L$ approximation, as in (\ref{Jmagnon}), we obtain
\beq
I_\phi[\bar{X},\bar{s};y]  = -\frac{g}{\pi^2}\, \sin{\frac{p}{2}} \;\frac{x_f^{4}}{ y^4\,(x_f-y)^4 }\,.
\eeq
Therefore, we conclude that the one-magnon  contribution to  the 3-point function is
\begin{equation} \label{3p-GM}
\langle {\cal O}_A(0) {\cal O}_A(x_f) {\cal L}(y) \rangle \approx - \frac{g}{\pi^2}\, \sin \frac{p}{2}  \;\frac{1}
{ x_f^{\,2\left(J+ 4g\sin \frac{p}{2} \right) -4}\,y^4\,(x_f-y)^4 }\,,
\end{equation}
which agrees with the expected result
\beq
2\pi^2  a_{{\cal L}AA} = -
g^2 \frac{\partial \Delta_A}{\partial g^2} \approx - 2 g  \sin \frac{p}{2} \,.
\eeq

\subsection{Spinning string on $AdS_5\times S_5$}

The examples considered in the previous three sections only dealt with relatively simple string configurations with particle-like motion in the $AdS_5$ part. We are interested in testing our approach in a more general setup, where the bulk-to-boundary propagator for the supergravity field varies along the 
string worldsheet for fixed worlsheet time $\tau$. We shall study the spinning string solution with angular momenta both in the $S^5$ and the $AdS_5$ factors \cite{Arutyunov:2003za}, whose 2-point function was also computed in \cite{Janik:2010gc}. The study of this solution is necessarily more intricate, because the way in which the string approaches the boundary depends non-trivially on the $AdS_5$ rotation. 
In this case it is convenient to use embedding coordinates.
Again, the starting point of this calculation is the Polyakov action in the conformal gauge, coupled to the metric and dilaton,
\begin{multline}
S_P[X,s,\Phi] = -g \int_{-s/2}^{s/2} d\tau \int_{0}^{2\pi} d\sigma\,e^{\phi/2}\,\Big[\eta^{\alpha\beta} \partial_{\alpha} Y^{a}\partial_{\beta} Y^{b}G_{ab}+\eta^{\alpha\beta}\partial_{\alpha} X^{i} \partial_{\beta} X^{j}\, G_{ij}+ \\ \tilde{\Lambda} \,(Y^2+1)+\Lambda\, (X^2-1)\Big]\,,\ \ \ \ 
\label{eq:embedding}
\end{multline}
where, as before, we have set the $AdS_5$ length and the radius of the $S^5$ to unity and $G$ is the embedding metric. The classical solution  representing a spinning string is given by
\begin{align}
& Y^0 = \frac{1}{2} \left[\cosh \rho _0 \left(\frac{R^2+1}{R}\, e^{\kappa  \tau }+\frac{e^{-\kappa  \tau }}{R}\right)+\frac{2 \sinh \rho _0 \cos (\tilde{\omega}\,\tau+\sigma )}{R}\right], \nonumber \\ 
& Y^2 + Y^0= \cosh \rho _0 \left(\frac{R}{2}+\frac{e^{\kappa  \tau }}{R}\right)\,,\qquad Y^3-Y^4 = e^{\kappa  \tau } \cosh \rho _0\,, \nonumber \\
& Y^4 = \sinh \rho _0 \cos (\tilde{\omega}\, \tau+\sigma)\,, \qquad Y^1 = Y^5 = 0\,,\qquad \tilde{\Lambda} = -\kappa^2\,, \nonumber\\
& X^1+i\, X^2 = e^{i\, (\omega\,\tau-\sigma)}\,, \qquad X^i = 0,\text{ for } i > 2\,, \qquad \Lambda = \omega^2 - 1\,,\label{eq:tute}
\end{align}
where $\tilde{\omega} = \sqrt{1-\kappa^2}$. The conserved charges of this solution can be readily obtained as functions of $\rho_0$, $\omega$ and $\kappa$, and are given by $J = 4\pi g\,\omega$, $S = 4\pi g\,\tilde{\omega}\,\sinh^2\rho_0$ and $E = 4\pi g\,\kappa \cosh^2\rho_0$, where $J$, $S$ and $E$ are the angular momentum on the $S^5$, angular momentum on the $AdS_5$ and energy, respectively. This solution has the required boundary conditions if we further identify $R = x_f$ and
\begin{equation}
s = \frac{2}{\kappa}\log \frac{x_f}{\epsilon}\,,
\end{equation}
where, as in \cite{Janik:2010gc}, we have conveniently absorbed a factor of $\cosh\rho_0$ in $\epsilon$.

In order to calculate the 2-point function, we have to apply the procedure described in \cite{Janik:2010gc}, which amounts to considering \reef{NewAction} instead of $S_P[\bar{X},s,\Phi = 0]$ defined in (\ref{eq:embedding}). Note, however, that the $AdS_5$ part in (\ref{eq:tute}) also depends on $\sigma$, which will lead to non-zero values of the zero-modes defined in (\ref{ZeroModes}). 
A computation shows that  \reef{NewAction} becomes
\begin{equation}
\tilde{S}_P[\bar{X},\kappa,\Phi = 0]= 4 \pi  g \left[\kappa-\frac{S \sqrt{1-\kappa ^2}}{2 \pi g\kappa} -\frac{1}{\kappa}\left(1+\frac{J^2}{16 \pi^2 g^2}\right)\right]\log\frac{x_f}{\epsilon}\,.
\end{equation}
It will be more convenient to evaluate the saddle point with respect to $\kappa$ instead of the modular parameter $s$. The saddle point in $\kappa$ yields the following condition
\begin{equation}
1+\bar{\kappa}^2+\frac{J^2}{16 \pi^2 g^2 }+\frac{S}{2 \pi g \sqrt{1-\bar{\kappa}^2} } = 0\,.
\label{eq:saddlerot}
\end{equation}
This condition is exactly the Virasoro constraint for the spinning string (\ref{eq:tute}). At the saddle point we recover the usual 2-point function behaviour
\begin{equation}
e^{\tilde{S}_P[\bar{X},\bar{\kappa},\Phi = 0]} = \left(\frac{\epsilon}{x_f}\right)^{8 \pi  g \left(\kappa _s+\frac{S \kappa _s}{4 \pi g \sqrt{\kappa _s^2+1}}\right)}\,,
\end{equation}
which gives the correct scaling dimension $\Delta_A = E = 4 \pi  g \left(\kappa _s+\frac{S \kappa _s}{4 \pi g \sqrt{\kappa _s^2+1}}\right)$, where we have defined $\bar{\kappa} = i \kappa_s$. Note that in the $S\to0$ limit we recover the circular spinning string of Section 4.2, as we should.

Now we will obtain the 3-point function. First we need to evaluate (\ref{I_integral}). In contrast to the other cases, the integrand in this case strongly depends on $\sigma$, and the integrals might appear to be very complicated. However, because we are integrating $\sigma$ from $0$ to $2\pi$ we can rewrite our integral as an integral in the complex plane by considering the complex variable ${\rm w} = e^{i\sigma}$. The integral can then be computed using the residues theorem. For fixed wordsheet time $\tau$, the integrand has two poles, one of which has zero residue and the other gives the relevant contribution. The intermediate steps are too cumbersome to be presented here, so we just state the final result
\begin{equation}
I_\phi[\bar{X},\bar{\kappa};y] = -\frac{\Delta_A}{4\pi^2 \kappa_s^2}
\left(1+\frac{S}{2\pi g \tilde{\omega}}\right)\frac{x_f^4}{y^4(x_f-y)^4}\,.
\end{equation}
Therefore, we conclude that, to leading order in $g$,
\beq
\langle{\cal O}_A(0) {\cal O}_A(x_f){\cal L}(y) \rangle \approx-\frac{g}{\pi} \;
\frac{1}{ x_f^{2 \Delta_A-4} \,y^4\,(x_f-y)^4 }\,.
\eeq
As happened in the previous cases, the spacetime dependence is the one required by conformal invariance. Moreover, to leading order in $g$, the coupling $a_{{\cal L}AA}$ determined in this way agrees with the RG expectation, since
\beq
2 \pi^2 a_{{\cal L}AA} = - g^2 \frac{\partial \Delta_A}{\partial g^2} \approx -\frac{g}{\pi} \,,
\eeq
where we kept the angular momenta $S$ and $J$ fixed when taking the derivative. We used (\ref{eq:saddlerot}) to determine $\partial_g \kappa_s$, which leads to $\kappa_s \approx 1 +O(1/g)$.

\section{Conclusion}

One of the present challenges in the gauge/gravity duality is to search for new techniques to compute correlation functions 
of single trace operators in ${\cal N}=4$ SYM, therefore leading to the exact solution of this theory possibly using integrability, 
at least in the planar limit. This paper is focused  on  such new search, both at weak and strong coupling. 

We started by presenting generic arguments based on the renormalization of operators when a CFT is deformed by a marginal or
irrelevant operator ${\cal D}$. These arguments were independent of the value of the coupling at the CFT fixed point (a fixed line in the case 
of ${\cal N}=4$ SYM). To linear order in the deformation parameter $u$, we can write the anomalous dimension matrix as
\beq
{\cal H}^u={\cal H} + u {\cal H}'\,.
\eeq
We showed that the matrix element ${\cal H}'_{AB}$ for two operators ${\cal O}_A$ and ${\cal O}_B$
is  determined by the coupling $a_{{\cal D}AB}$. 
For  ${\cal N}=4$ SYM and at weak 't Hooft coupling, we can start by 
diagonalizing ${\cal H}$ using integrability to some order in the coupling. Then the problem of computing $a_{{\cal D}AB}$
amounts to determining the matrix elements ${\cal H}'_{AB}$. We saw how to implement these ideas for the simplest case of the
exact Lagrangian deformation, where the action of the matrix  ${\cal H}'({\cal L})$ on the basis of operators represented as a spin chain 
is known (it is just the derivative with respect to the coupling of the anomalous dimension matrix at the critical point). 
In this case one needs to compute the matrix elements of  ${\cal H}'({\cal L})$ between Bethe roots. A very interesting open problem 
is to extend this procedure to other deformations, therefore allowing for a systematic computation of the couplings $a_{{\cal D}AB}$
in perturbation theory using integrability. When ${\cal D}$ is a chiral operator  in the same multiplet of the Lagrangian, 
for example the energy-momentum tensor, we expect that the action of ${\cal H}'({\cal D})$ on the basis represented as a spin chain 
can be obtain acting with the supersymmetry algebra on ${\cal H}'({\cal L})$. 
It would be very interesting to see how far one can go with this type of approach.

The gauge/gravity duality can be used to compute ${\cal N}=4$ SYM correlation functions at strong coupling, but in practice one is limited to
the supergravity approximation which only includes chiral operators.
We have improved on this limitation, by  including two insertions of an operator dual to a heavy string state, and then studied the
case of 3-point correlation functions with one extra chiral operator.
We considered specific examples with the Lagrangian operator, checking agreement with the expected result from renormalization group arguments,
since in this case the coupling $a_{{\cal L}AA}$ is simply related to the derivative with respect to the 't Hooft coupling of $\Delta_A$. 
This is an important check, since it gives confidence that the method can be applied to other chiral operators. 
Our computation of the string theory partition function is based on a saddle point approximation to the string path integral that describes the propagation
of the heavy state, generalising
the analysis of the 2-point function introduced in   \cite{Janik:2010gc}.  Other chiral operators can be included because 
the heavy string acts as a tadpole for the supergravity fields, which may then propagate to the boundary of $AdS$ if sources are present therein.
An alternative way to think of this computation is to realise  that the  supergravity fields act as sources in the equations of motion for the worldsheet
fields of
the heavy string worldsheet, therefore deforming it.
In fact, the same occurs in the case of three large operators, whose 3-point function at strong coupling ought to be determined by a string worldsheet
with fixed boundary condition on three points in the boundary of $AdS$.  In \cite{Janik:2010gc} this was shown to yield the correct 
conformal dependence of the 3-point function on the boundary points, but the evaluation of the coupling $a_{ABC}$, related to the area of
the string worlsheet minimal surface, is still an open problem. It is expected that such computation will 
show a direct relation with integrability, but that remains to be seen.



\newpage

\begin{center} 
{\bf Acknowledgements} 
\end{center}
The authors are grateful to Jo\~ao Penedones and Pedro Vieira for helpful discussions. We also thank Konstantin Zarembo for sharing with us a draft of his work \cite{Zarembo}, which has some overlap with Section 4 of this article. M.S.C. wishes to thank NORDITA for hospitality during the  programme 
{\em Integrability in String and Gauge Theories; AdS/CFT Duality and its Applications}.
D.Z. is funded by the FCT fellowship SFRH/BPD/62888/2009. This work was partially funded by the research 
grants PTDC/FIS/099293/2008 and CERN/FP/109306/2009.
\emph{Centro de F\'{i}sica do Porto} is partially funded by FCT.

\section*{Appendix A: Irrelevant deformations}

In this appendix we re-write the formulae of Section 2 for the case of irrelevant deformations. The main results for the renormalization
of operators are essentially the same, so we shall be brief. The first modification is that we must be careful with the running of the coupling $u$,
so that correlation functions involving a renormalized operator  of the undeformed theory  ${\cal O}_A$ become
\begin{equation}
\left\langle {\cal O}_A(x) \cdots \right\rangle_u
 = \left\langle {\cal O}_A(x) \cdots \right\rangle -  u\, \Lambda^{4-\Delta} \int d^4y \left\langle {\cal O}_A(x)  \,{\cal D}(y)\cdots \right\rangle \,,
\end{equation}
where we work to linear order in the deformation parameter $u$, as defined at the cut-off scale $\Lambda$.
Using the OPE expansion between ${\cal D}$ and ${\cal O}_A$ given in (\ref{OPE}), we conclude that the 
region of integration $y\sim x$ contributes with
\begin{equation}
\Lambda^{4-\Delta}  \int \frac{d^4y}{|x-y|^{\Delta+ \Delta_A-\Delta_B}} 
\approx
\frac{ 2\pi^2\Lambda^{ \Delta_A-\Delta_B}}{\Delta +\Delta_A-\Delta_B-4}  \,,
\end{equation}
for $\Delta_B \le \Delta_A$. We renormalize the operators of the deformed theory according to
\begin{equation}
{\cal O}^u_A = {\cal O}_A + 
u \sum_{\Delta_B\le\Delta_A}  2\pi^2a_{{\cal D}AB}  \,\frac{\Lambda^{ \Delta_A-\Delta_B} }{\Delta +\Delta_A-\Delta_B-4} \,{\cal O}_B\,,
\label{Irr_renormalized}
\end{equation}
so that the  correlation functions $\left\langle {\cal O}^u_A(x) \cdots \right\rangle_u$ are finite. With this prescription the
two-point function between operators $\left\langle {\cal O}^u_A(x)  {\cal O}^u_B(0) \right\rangle_u$ remains the same.
This is expected because the deformation is irrelevant and therefore, to leading order in $u$, it is not expected to change the anomalous dimension
of operators. We remark that the constant renormalization in (\ref{Irr_renormalized}) for $\Delta_A=\Delta_B$ guarantees that this two-point function
remains diagonal and with the same normalization.

The above renormalization scheme corresponds to a 
renormalization matrix (\ref{RMatrix}) with entries
\beq
Z_{AB} = \delta_{AB} + 
u \,2\pi^2 a_{{\cal D}AB}\, \frac{ \Lambda^{ \Delta_A-\Delta_B}}{\Delta + \Delta_A-\Delta_B-4}  \,.
\eeq
The corresponding anomalous dimension matrix $\Gamma_{AB}$ defined in
(\ref{Gamma}) is the same  as for the marginal case given in 
(\ref{DeformedGamma}). The Callan-Symanzik equation will now include the 
running of the coupling $u$. For example, for 2-point functions we have
\begin{equation}
\left( \frac{\partial \  \ }{\partial\ln\Lambda} + \beta_u\, \frac{\partial\ }{\partial u} \right)
\left\langle  {\cal O}_A(x) {\cal O}_B(0) \right\rangle_u +
\sum_I \Gamma_{AI}  \langle  {\cal O}_I(x) {\cal O}_B(0) \rangle_u +
\sum_I \Gamma_{BI}  \langle  {\cal O}_A(x) {\cal O}_I(0) \rangle_u =0\,,
\end{equation}
with $\beta_u= (\Delta - 4) u$. This equation is verified provided (\ref{DeformedGamma}) holds.

Finally, it is useful to define the renormalized operators starting from the undeformed bare theory. In this case the renormalization matrix
${\cal Z}_{AB}$ has entries
\beq
{\cal Z}_{AB} = \delta_{AB} \,\gamma_A+ 
u \,2\pi^2 a_{{\cal D}AB}\, 
\frac{ \Lambda^{ \Delta^0_A -\Delta^0_B + \gamma_A }}{\Delta + \Delta_A-\Delta_B-4}  \,.
\eeq
The anomalous dimension matrix again has entries given by (\ref{DeformedH}). The relation between the entries of the deformed anomalous dimension 
matrix ${\cal H}'$, computed in a diagonal basis of the anomalous dimension matrix of the critical  theory,  and the couplings $a_{{\cal D}AB}$
is as for the marginal case presented at the end of Section 2.

\end{document}